\documentclass[%
 reprint,
 superscriptaddress,
 amsmath,amssymb,
 aps,
]{revtex4-1}

\usepackage{graphicx}
\usepackage{dcolumn}
\usepackage{bm}
\usepackage{tikz} 
\usetikzlibrary{automata,positioning}

\let\oldhat\hat

\renewcommand{\hat}[1]{\oldhat{\mathbf{#1}}}

\begin{document}


\title{Signatures of motor susceptibility in the dynamics of a tracer particle in an active gel}

\author{Nitzan Razin}
 \affiliation{Department of Chemical and Biological Physics, Weizmann Institute of Science, Rehovot 76100, Israel}
\author{Raphael Voituriez}%
\affiliation{Laboratoire Jean Perrin and Laboratoire de Physique Th\'eorique de la Mati\`ere Condens\'ee, CNRS / Sorbonne Universite, 75005 Paris, France}%

\author{Nir S. Gov}
 \affiliation{Department of Chemical and Biological Physics, Weizmann Institute of Science, Rehovot 76100, Israel}
\date{\today}

\begin{abstract}
We study a model for the motion of a tracer particle inside an active gel, exposing the properties of the van Hove distribution of the particle displacements. Active events of a typical force magnitude give rise to non-Gaussian distributions, having exponential tails or side-peaks. The side-peaks appear when the local bulk elasticity of the gel is large enough and few active sources are dominant. We explain the regimes of the different distributions, and study the structure of the peaks for active sources that are susceptible to the elastic stress that they cause inside the gel. We show how the van Hove distribution is altered by both the duty cycle of the active sources and their susceptibility, and suggest it as a sensitive probe to analyze microrheology data in active systems with restoring elastic forces.
\end{abstract}

\maketitle

\emph{Introduction ---} Active gels, composed of a cross-linked network of biopolymers that contains molecular motors which produce active forces, are in-vitro model systems for exploring non-equilibrium physics as well as studying the dynamics inside living cells \cite{Backouche2006,Mizuno2007,Cates2008,MacKintosh2008,MacKintosh2010,Toyota2011,Stuhrmann2012,Silva2014,Sonn-Segev2017,Sonn-Segev2017JPCM,Alvarado2013,Marchetti2013RMP,BenIsaac2015}. The dynamics of an active gel can be probed experimentally by following the displacements of a passive tracer particle \cite{Toyota2011,Stuhrmann2012,Silva2014,Sonn-Segev2017}. The time-dependent distribution of the displacements, the van Hove distribution (VHD), was observed to be an indicator of activity inside the gel, exhibiting distinct non-Gaussian form in the presence of activity \cite{Toyota2011,Stuhrmann2012,Silva2014}. It was further predicted theoretically that when the motion is dominated by few sources of active forces, and the elastic stiffness of the gel is high, there should appear side-peaks in the VHD \cite{BenIsaac2015}. This peak structure arises from the discrete displacement of the tracer particle due to the force balance between the active force and the elastic restoring force.

In this paper, we study how the VHD changes with the properties of the active forces.
We show that the model introduced in \cite{BenIsaac2015} has parameter regimes where the VHD has exponential tails, similar to those commonly measured for tracers in active gels \cite{Toyota2011, Stuhrmann2012,Silva2014}.
We then extend the model in \cite{BenIsaac2015} to account for susceptibility of the active event sources to the elastic stress in the active gel. This is motivated by the known susceptibility of molecular motors, such as myosin II, to applied forces, which modify their kinetics \cite{Klumpp2005,Guerin2010,Wang2011,Wang2012,Ucar2017,Sung2015}. 
We show how the model parameters affect the peak structure in the VHD, thereby demonstrating that it is a sensitive probe for the properties of the active sources inside active gels and living cells. Our results may be relevant to many active systems with restoring elastic forces, such as artificial active membranes \cite{Gov2004,Girard2005,Turlier2018}, cellular membranes \cite{Ben-Isaac2011}, active polymers \cite{Loi2011,Ghosh2014,Weber2015}, particles within optical traps \cite{Gomez-Solano2010}, dense active fluids \cite{Berthier2002,Loi2008,Loi2011SM} and active fluctuations observed in the chromatin in the nucleus of living cells \cite{Malte2000,Weber2012,Bruinsma2014,Zidovska2015,Zidovska2017}.

\emph{Model definition ---} We start by studying a one dimensional model for the motion of a tracer particle in the active gel, which was introduced in \cite{BenIsaac2015}. The particle is trapped in a harmonic potential $U(x)=\frac{1}{2}kx^2$, representing the bulk elasticity of the gel. This description is valid at short times, in which the particle is trapped, since at long times the actin network reorganizes and the particle performs free diffusion \cite{Toyota2011,Sonn-Segev2017}. In addition, the particle moves due to the force applied by $N$ "motors", each representing a source of active events which applies a force with typical magnitude $F_0$ for an exponentially distributed time duration with average $k_{\textrm{off}}^{-1}$. At first we assume, as in \cite{BenIsaac2015}, that the motors are independent and identically distributed: each of them can be either on or off. When a motor is off, it doesn't apply any force to the particle. The motor turns on after an exponentially distributed duration, with rate $k_{\textrm{on}}$. Once the motor is on, a direction (left or right) is randomly chosen and the motor applies a force of magnitude $F_0$ in that direction to the particle.

The Langevin equation of motion for the particle position $x$ is thus:
\begin{equation} \label{eq:LangevinEOM}
m\ddot{x} = -\gamma\dot{x} -kx - \sum\nolimits_{i=1}^N M_i(t) F_0 + \xi(t)
\end{equation}
Where $m$ is the particle mass, $\gamma$ is a friction coefficient, $M_i$ is a variable representing the state of the i-th motor, and is equal to $0$ when the motor is off and $\pm1$ when the motor is on in the right/left direction. $\xi(t)$ is white noise that can be of thermal or other origin (such as activity from many sources that are far from the particle). We neglect $\xi$, since small white noise does not significantly affect the properties of the VHD peak structure in which we are interested. Moreover, in cellular systems active temperatures are usually much larger than thermal ones.
The motion of tracer particles in cellular systems and artificial active gels is typically overdamped. Accordingly, our model focuses on this regime, where the friction timescale $m/\gamma$ is smaller than the other timescales, $k_{\textrm{off}}^{-1}$, $k_{\textrm{on}}^{-1}$ and $\gamma/k$. Yet we keep the general form of the Langevin equation in all of our simulations for accuracy and consistency, since as the susceptibility of the motors to the external force increases, the rates of motor state change can become greater than the friction timescale $m/\gamma$, crossing into the underdamped regime.

In the following, we focus on the properties of a commonly measured characteristic of the motion of tracer particles in experimental systems \cite{Toyota2011,Stuhrmann2012,Silva2014} - the van Hove correlation function $P(\Delta x(\Delta t))$. It is defined as the distribution of particle displacements $\Delta x \equiv x(t+\Delta t)-x(t)$ over a lag time $\Delta t$.

\emph{Qualitative behavior of the van Hove distribution ---} Two qualitative behaviors of the VHD of tracer particle displacements have been observed in active gels: Gaussian \cite{Sonn-Segev2017} and exponential tails \cite{Toyota2011, Stuhrmann2012,Silva2014}. Both behaviors can be explained within the model defined above, in different regimes of the bulk elasticity, the number of motors and the lag time $\Delta t$.

If a tracer particle is far enough from all active motors, its motion is caused by thermal noise combined with the small effect of each of the distant motion of many motors. In this case, its VHD will be Gaussian, irrespective of the value of $k$ \cite{BenIsaac2015}.

Exponential tails, as were observed in active gels \cite{Toyota2011,Stuhrmann2012,Silva2014} and living cells \cite{Bursac2005,Fodor2015,Fodor2018,Ahmed2018}, can result from forces applied by few nearby motors (as suggested in \cite{Toyota2011}), if the local bulk elasticity is small. In the Supplemental Information Section 1 \cite{Supplemental}, we show that within our model, in the limit of small bulk elasticity and small duty cycle (ratio of time in which a motor is on) $p_{\textrm{on}}\equiv\frac{k_{\textrm{on}}}{k_{\textrm{on}}+k_{\textrm{off}}}$, the exponential statistics of the  motor active burst duration results in exponential tails of the VHD. 
Thus our model is consistent with observations of exponential tails in active gels.

\begin{figure}[h]
\includegraphics[width=1\linewidth]{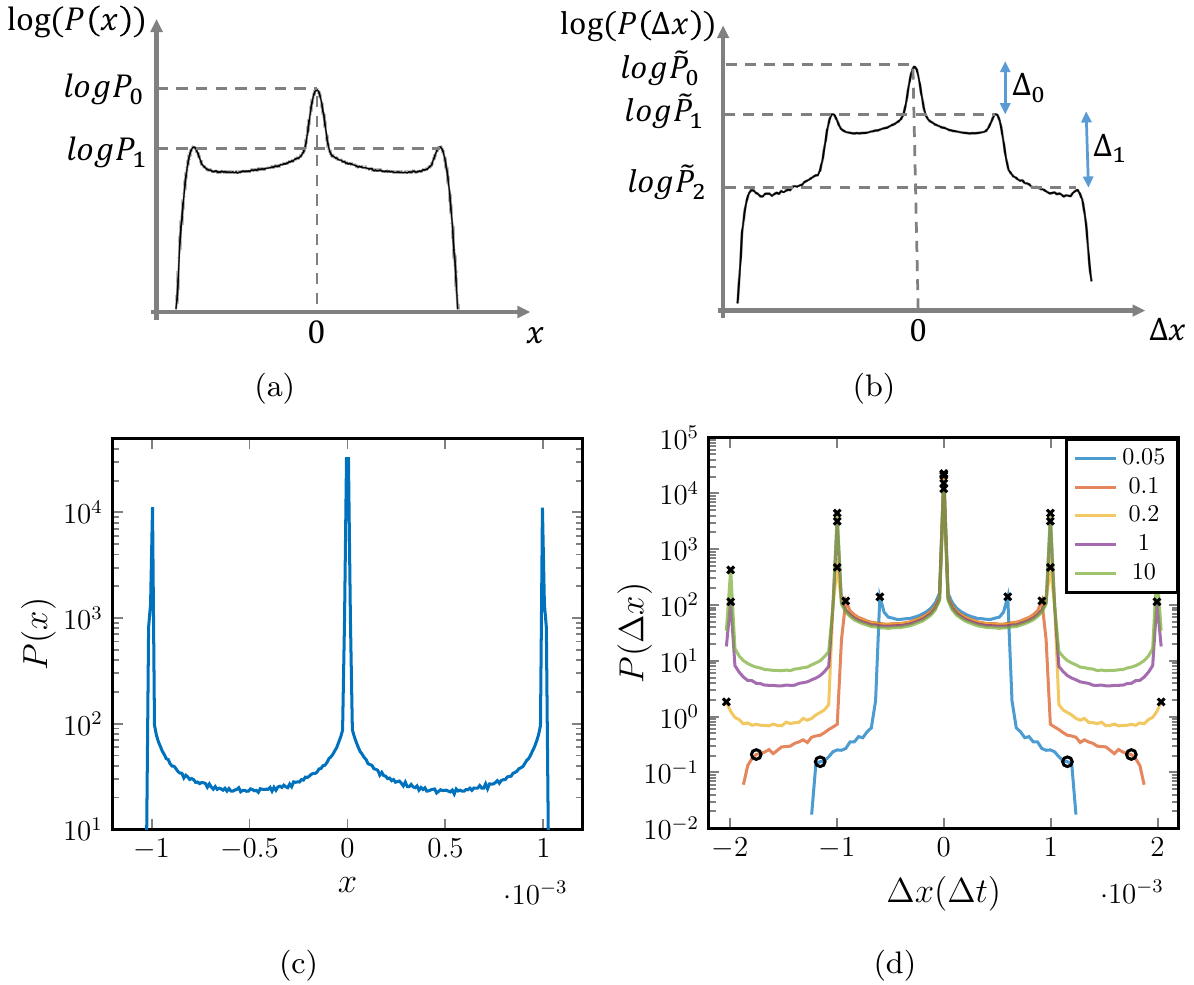}
  \caption[]{(a) Sketch of the steady state position distribution of the particle for $N=1$ motor, with the peak notations marked. (b) Sketch of the matching long time VHD of particle displacements  ($P(\Delta x(\Delta t\rightarrow\infty))$). Notation for the distribution peaks and consecutive peak differences is marked. (c) The steady state particle position distribution for $N=1$ adamant motor ($k_{\textrm{on}}=0.43$, $k_{\textrm{off}}=1$, $k=1000$, $F_0=1$, $\gamma=50$). (d) The VHD for the same system as in (c). Different colors represent different lag times $\Delta t$. Detected peaks are marked by black x's, and shoulders by black circles.}
\label{fig:dist_vH_sketch}
\end{figure}

A third behavior was predicted theoretically in \cite{BenIsaac2015} to occur within an experimentally relevant regime: if a tracer particle is in an environment with large bulk elasticity, and in the vicinity of only a few motors, then side peaks appear in the long-time VHD (Fig.~\ref{fig:dist_vH_sketch}a,b), while at shorter times, these peaks become shoulders (Fig.~\ref{fig:dist_vH_sketch}c,d). We now investigate the dependence of this peak structure on the properties of the active sources.

\emph{van Hove distribution peak ratio ---} In the strong confinement limit $k_{\textrm{off}}, k_{\textrm{on}}\ll k/\gamma$, the steady state particle position distribution $P(x)$ has sharp peaks located at positions where the motor force is balanced by the harmonic potential force \cite{BenIsaac2015}. The particle spends most of the time at the peaks of $P(x)$: the time it takes the particle to move the distance between adjacent peaks $x_0 \equiv F_0/k$ is $\gamma/k$ (as calculated in the SI Section 2 \cite{Supplemental}), and in this limit it is much smaller than the typical times for a motor state change - $k_{\textrm{off}}^{-1}$ and $k_{\textrm{on}}^{-1}$.
$P(x)$ then has one central peak at $x=0$, and $2N$ side peaks, $N$ on each side of $x=0$ (Fig.~\ref{fig:dist_vH_sketch}a).
The long-time VHD is simply a self-convolution of $P(x)$: $P(\Delta x(\Delta t \to \infty))=P(x)*P(x)$. Thus it has one central peak and $4N$ side peaks. Denote the value of $P(\Delta x)$ in each of the peaks $\tilde{P}_i$, where $i=-2N,-2N+1,...,2N$. In order to quantify the peaks structure of the VHD, we measure the ratio of consecutive peak height differences in log scale (Fig.~\ref{fig:dist_vH_sketch}b). The system is reflection-symmetric and therefore $\tilde{P}_{-i}=\tilde{P}_{i}$. Thus for convenience we can focus without loss of generality on the peaks with non-negative indexes, which are at non-negative $x$ values. We define the consecutive peak height differences in log scale: $\Delta_i = \log(\tilde{P}_{i}/\tilde{P}_{i+1})$. We study the dependence of the ratio of two consecutive log peak height differences: $r_i(\Delta t) = \Delta_i/\Delta_{i+1}$ on our model parameters, and show that its value can be a signature of motor susceptibility to external force.

\begin{figure}[t]
\includegraphics[width=1\linewidth]{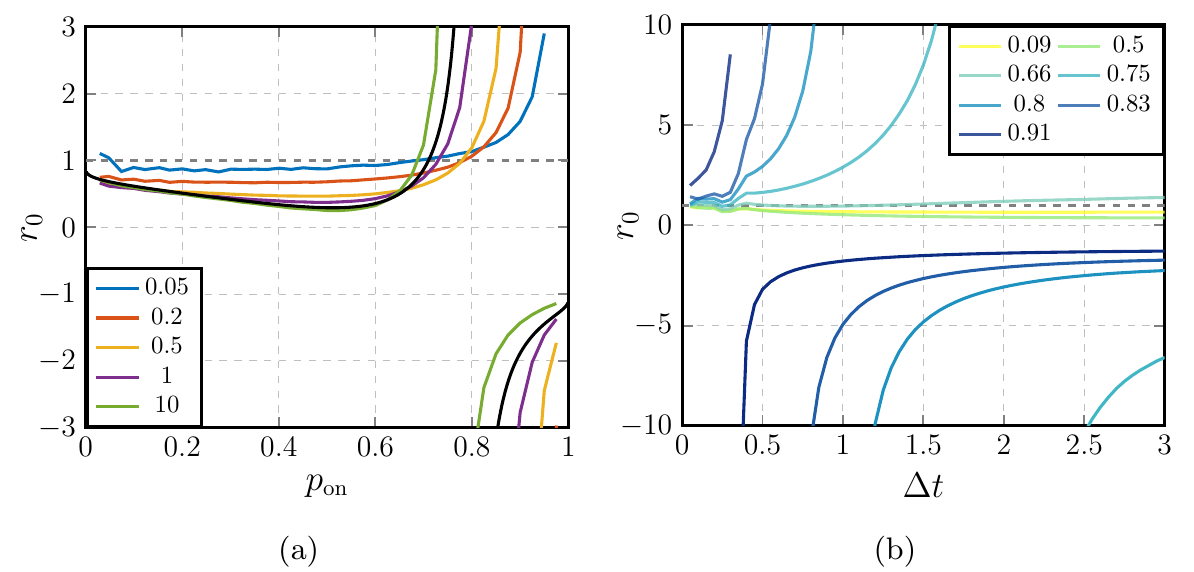}
  \caption[]{$r_0$ in simulations with $N=1$ adamant motor, $k_{\textrm{off}}=1$ and varying $k_{\textrm{on}}$ ($k=1000$, $F_0=1$, $\gamma=50$): (a) $r_0$ vs $p_{\textrm{on}}$. Different colors represent different lag times $\Delta t$. The black line is the approximate theoretical result for the fast particle limit at long lag times. (b) $r_0$ vs $\Delta t$. Colors represent different $p_{\textrm{on}}$ values. For $p_{\textrm{on}}\lesssim0.6$, $r_0<1$ at all lag times.}
\label{fig:N_1_adamant_r0}
\end{figure}

We begin by analyzing adamant motors, where the dynamics of the active sources do not depend on the stress in the gel, and the rates $k_{\textrm{on}}$, $k_{\textrm{off}}$ are constant.
In the SI Section 3 \cite{Supplemental}, we show that when $p_{\textrm{on}}\to0$, the long-time VHD has $r_i\to 1$. Furthermore, $r_i<1$ and quite close to 1 for all $p_{\textrm{on}}<p_{\textrm{on}}^c$ for $p_{\textrm{on}}^c \approx 0.6$ (Fig.~\ref{fig:N_1_adamant_r0}a, S2). Thus $p_{\textrm{on}}$ needs to be larger than $p_{\textrm{on}}^c$ in order to obtain a $r_i>1$. Moreover, numerical simulations (see SI Section 6 \cite{Supplemental}) show that this remains true for small $\Delta t$ values (Fig.~\ref{fig:N_1_adamant_r0}), in a sense that is clarified below.
As $\Delta t$ becomes smaller, the VHD becomes narrower, as the displacements are smaller over shorter times. Eventually, peaks (maximum points of $P(\Delta x)$) become shoulders
(Fig.~\ref{fig:dist_vH_sketch}d). A shoulder is a region of decreased slope, between two regions with a larger slope. We define the location of the shoulder to be the minimum point of $|P'|$.
As $\Delta t$ decreases, $P(\Delta x)$ continuously deforms until there is no longer a maximum point where $P'=0$, and instead $|P'|>0$ throughout the region. This happens since the tracer does not have time to reach the force balance position, and remain there, within time $\Delta t$.
When a peak $\tilde{P}_i$ becomes a shoulder as $\Delta t$ decreases, we continue to use the notation $\tilde{P}_i$ for the shoulder point height.


\emph{Susceptible motors ---} Next, we study the model when the active sources are susceptible to external force. While the susceptibility of molecular motors, which are the source of activity in active gels, is a known phenomenon \cite{Wang2011,Wang2012,Ucar2017,Klumpp2005,Sung2015,Guerin2010}, it is not obvious how to take it into account in our model due to several reasons. First, because the precise nature of the active events that affect the tracer particle position in the active gel is unknown \cite{Toyota2011,Sonn-Segev2017}. Second, because molecular motors respond to the force which acts on them locally, and it is unclear what this force is within our model, which does not assign a spatial position to the motors. We make the simple assumption that each of the motors is susceptible to the elastic force acting on the tracer particle, since this force is an indicator of the local stress in the gel.
This means that the stochastic dynamics of the motor state variables $M_i$ are now coupled to the particle position $x$, i.e. the motor state transition rates are functions of $x$. We therefore define $k_{\textrm{on}}^{\pm}(x)$ to be the rate of transition from the off state to the on state in the $\pm$ direction, and $k_{\textrm{off}}^{\pm}(x)$ to be the rate of transition from the on state in the $\pm$ direction to the off state.
In general, solving the coupled dynamics of the particle position and motor states to find $P(x)$ is difficult. We can solve it approximately in the fast particle limit. Let us consider the case of a single motor, $N=1$. Then, as shown in the SI Section 4 \cite{Supplemental}, the steady state particle position satisfies $P(x_0)=P(-x_0)=\frac{k_{\textrm{on}}(0)}{2 k_{\textrm{off}}(x_0)}P(0)$, where we denote $k_{\textrm{on}}(x) = k_{\textrm{on}}^{+}(x)+k_{\textrm{on}}^{-}(x)$ and $k_{\textrm{off}}(x_0) = k_{\textrm{off}}^+(x_0) = k_{\textrm{off}}^-(-x_0)$. The dimensionless ratio $k_{\textrm{on}}(0)/k_{\textrm{off}}(x_0)$ controls the steady state distribution. Thus the particle position distribution and therefore the long-time VHD are approximately affected only by tuning the susceptibility ($x$ dependence) of $k_{\textrm{off}}$, and indifferent to the dependence of $k_{\textrm{on}}$ on the external force (since at $x=0$ the external force is zero).
At shorter times 
this is not true since events where the motor turns on before reaching $x=0$, with $k_{\textrm{on}}$ that is dependent on the location $x$, can become dominant. For example, for a $\Delta t$ close to the time it takes the particle to move from $x_0$ to $-x_0$ with $M_1=-1$, $P(\Delta x=2x_0)$ is a result of events where the motor state changed from $+1$ to $0$ and nearly immediately to $-1$. The probability of such events depends on $k_{\textrm{on}}^{\pm}(x)$.

\begin{figure}[t]
\includegraphics[width=1\linewidth]{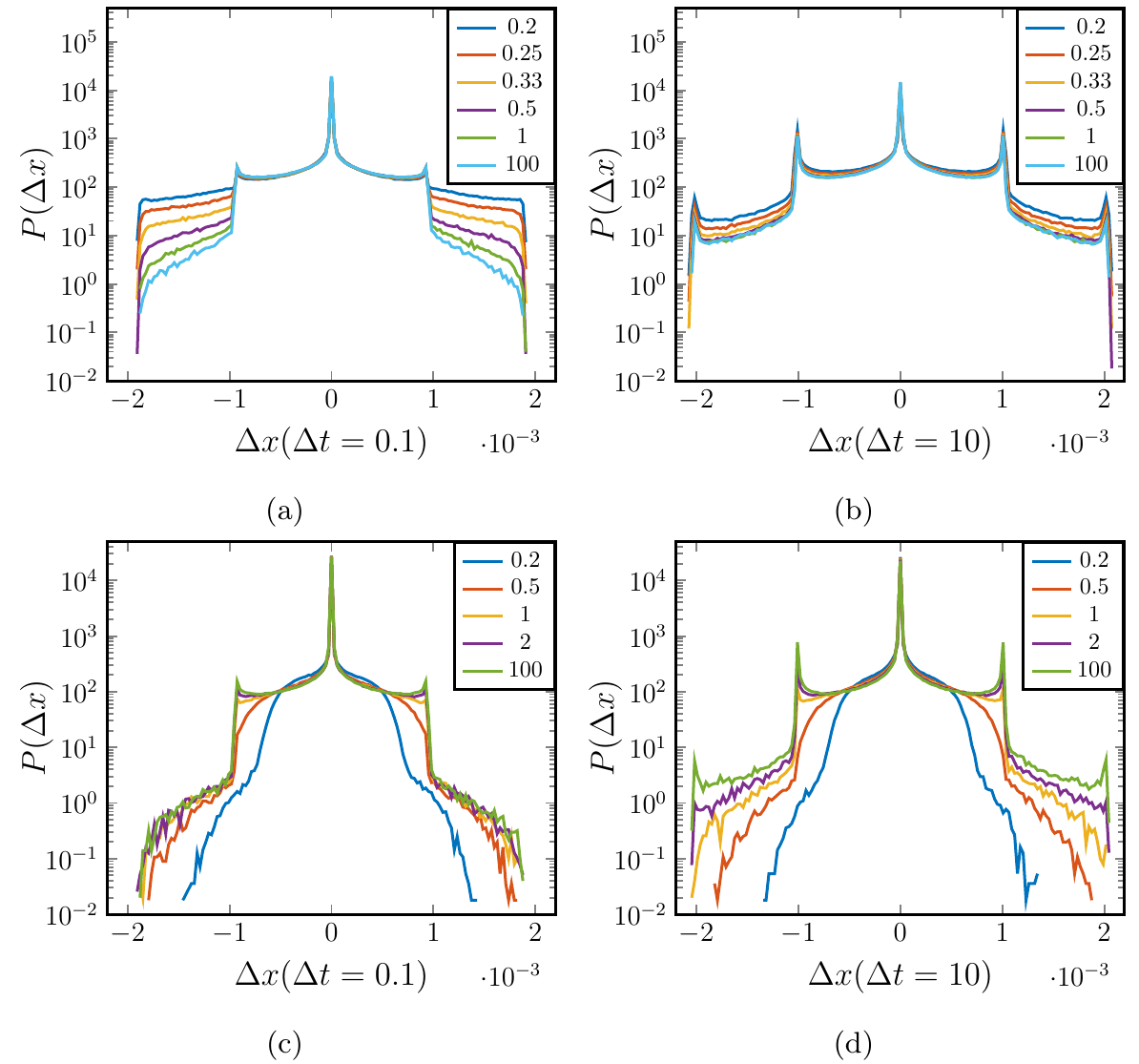}
  \caption[]{Comparison of systems of $N=1$ susceptible motor with susceptibility in the on or off rate. The VHD for various $F_1$ values is plotted for: (a-b) $k_{\textrm{on}}^{\pm}=e^{\mp kx/F_1}$, $k_{\textrm{off}}=10$, for lag times of $\Delta t=0.1$ (a) and $\Delta t=10$ (b). For $\Delta t=0.1$, increasing the susceptibility (decreasing $F_1$) increases the height of the shoulder $\tilde{P}_2$ and therefore increases $r_0$. (c-d) $k_{\textrm{on}} = 1$, $k_{\textrm{off}} = 10 e^{-Mkx/F1}$, for lag times of $\Delta t=0.1$ (c) and $\Delta t=10$ (d). Increasing the susceptibility causes peaks to become shoulders and move to smaller $|x|$. It does not increase $\tilde{P}_2$ or $r_0$. ($k=1000$, $F_0=1$, $\gamma=50$)}
\label{fig:N_1_k_on_vs_k_off_x_exp}
\end{figure}

\begin{figure}[t]
\includegraphics[width=1\linewidth]{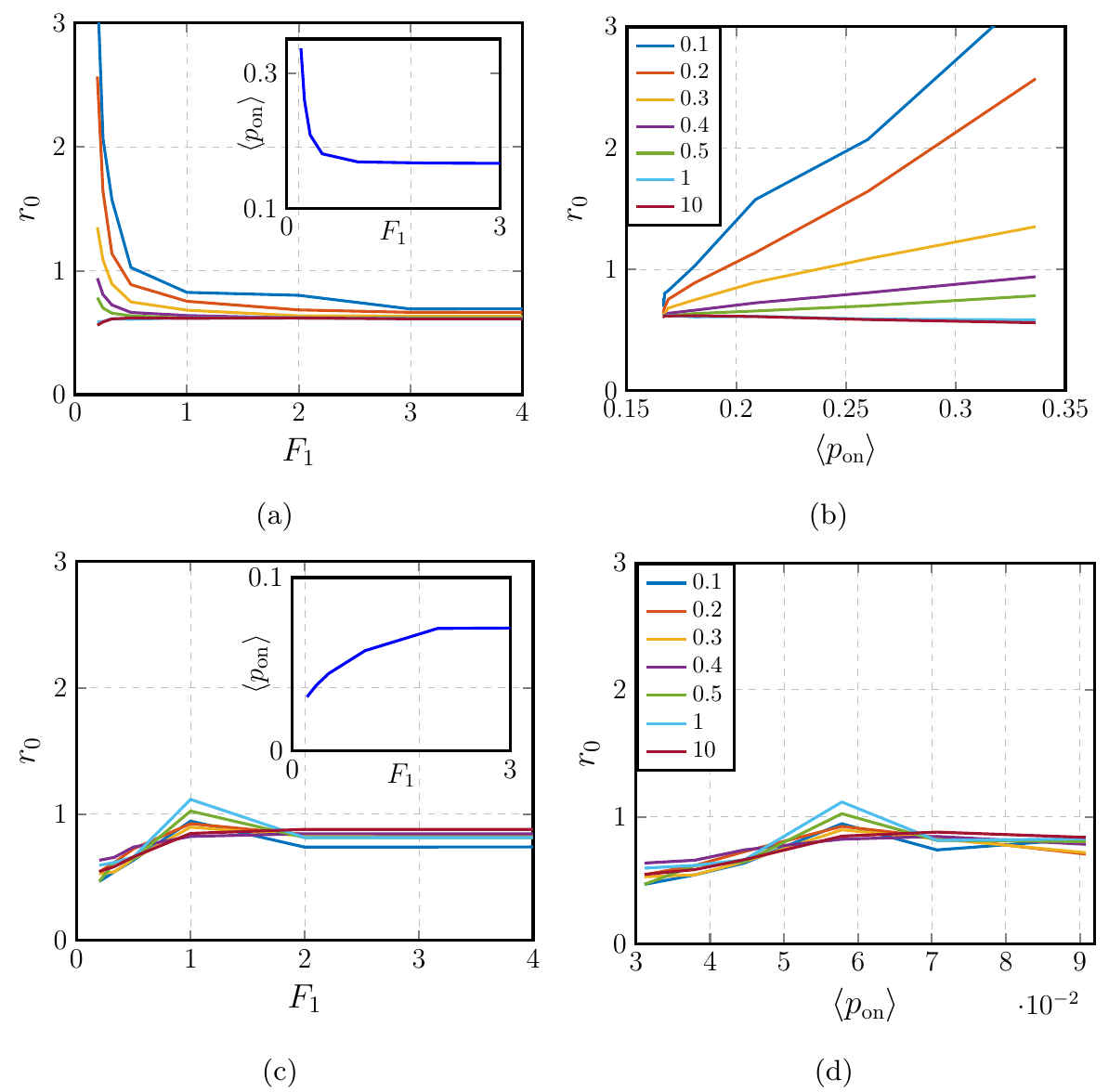}
  \caption[]{The effect of susceptibility on $r_0$. For $k_{\textrm{on}}^{\pm}=e^{\mp kx/F_1}$, $k_{\textrm{off}}=10$, $N=1$: (a) $r_0$ as a function of $F_1$ for various lag times $\Delta t$. Inset: The average duty cycle $\langle p_{\textrm{on}} \rangle$ as a function of $F_1$. (b) $r_0$ as a function of the average duty cycle $\langle p_{\textrm{on}} \rangle(F_1)$ for various lag times $\Delta t$. For a small enough $\Delta t$, $r_0>1$ for much smaller $p_{\textrm{on}}$ than in the adamant motor system. (c,d) Same as (a) and (b) for $k_{\textrm{on}} = 1$, $k_{\textrm{off}} = 10 e^{-Mkx/F1}$ ($k=1000$, $F_0=1$, $\gamma=50$)}
\label{fig:N_1_susc_r0}
\end{figure}

The simplest possibilities are that either the active event rate $k^{\pm}_{\textrm{on}}$ is susceptible to external force and the average event duration $1/k^{\pm}_{\textrm{off}}$ is constant, or vice versa. We considered both of these options, with the assumption that the dependence of the rates on the elastic force is exponential and such that the motor force tends to align with the elastic force. We define the two models we study as follows:

1. $k_{\textrm{off}}=\textrm{const}$, $k_{\textrm{on}}^{\pm}=k_{\textrm{on}}^{0}e^{\mp kx/F_1}$, where $k_{\textrm{on}}^0$ is the basal on rate in the absence of forces, and $F_1$ is a force scale that determines the sensitivity of the motors to external force. When $F_1 \to \infty$, the motors are adamant: $k_{\textrm{on}}^{+}=k_{\textrm{on}}^{-}=k_{\textrm{on}}^{0}$, and each motor is equally likely to push to the left or to the right, regardless of the elastic force on the particle.
In the infinite susceptibility limit $F_1 \to 0$, for $x>0$: 
$k_{\textrm{on}}^{+} \to 0$, $k_{\textrm{on}}^{-} \to \infty$ and the motor turns on in the direction of the elastic force infinitely fast.

2. $k_{\textrm{on}} = \textrm{const}$, $k_{\textrm{off}} = k_{\textrm{off}}^{0} e^{-Mkx/F1}$.
In the adamant motor limit $F_1 \to \infty$:
$k_{\textrm{off}}^{+}=k_{\textrm{off}}^{-}=k_{\textrm{off}}^{0}$, and the motor is equally likely to turn off whether it is pushing with or against the elastic force.
In the infinite susceptibility limit $F_1 \to 0$, for $x>0$: 
$k_{\textrm{off}}^{-} \to 0$, $k_{\textrm{off}}^{+} \to \infty$, and the motor immediately turns off if it is pushing against the elastic force, and never turns off if it is pushing with the elastic force.

Since in experiments using acto-myosin active gels, the duty cycle of the active sources is likely small \cite{Sonn-Segev2017,Mizuno2007, Toyota2011,BenIsaac2015}, we would like to explore the peaks structure obtained in this regime. Specifically, since in the small $p_{\textrm{on}}$ regime we find $r_0<1$ for adamant motors, we would like to find if susceptibility of the active sources can result in $r_0>1$ even at a small average duty cycle (note that for susceptible motors the average duty cycle is a result of the system dynamics).

From numerical simulations, we find that a susceptible $k_{\textrm{off}}$ does not increase $r_0$ above 1 (For N=1 see Fig.~\ref{fig:N_1_k_on_vs_k_off_x_exp}c,d, Fig.~\ref{fig:N_1_susc_r0}c,d, for N=2 see Fig.~S3c,d, Fig.~S4c,d \cite{Supplemental}). On the other hand, a susceptible $k_{\textrm{on}}$ causes $r_0$ to be greater than 1 at smaller $p_{\textrm{on}}$ values than the one for adamant motors (For N=1 see Fig.~\ref{fig:N_1_k_on_vs_k_off_x_exp}a,b, Fig.~\ref{fig:N_1_susc_r0}a,b, for N=2 see Fig.~S3a,b, Fig.~S4a,b \cite{Supplemental}).

\emph{Discussion ---} We considered a model for the motion of a tracer particle in an elastic gel due to active force sources of a typical force magnitude. We find that the two experimentally observed behaviors of the VHD of the tracer position displacements in actomyosin gels: Gaussian \cite{Sonn-Segev2017} and exponential tails \cite{Toyota2011, Stuhrmann2012,Silva2014} can be obtained in our model for different parameter values.
Additionally, we explored the structure of side-peaks in the VHD in the limit of strong confinement ("fast particle") $k_{\textrm{off}}, k_{\textrm{on}}\ll k/\gamma$. We characterized the peak structure using the difference between adjacent peaks. We showed that for adamant motors, the ratio of the first two differences $r_0$ is larger than $1$ only for large $p_{\textrm{on}}$ values. When the active event rate is susceptible to the elastic stresses in the gel (which are quantified by the force on the tracer), such that the motor and elastic force tend to align, we find that $r_0>1$ is obtained even when the average duty cycle is small. When the active event duration is susceptible to the elastic stress, $r_0$ is not increased by the susceptibility.
Thus, if the average duty cycle of the active events is small, the peak and shoulder structure can be an indication for susceptibility to external force of the active event rate. This information can shed light on the microscopic nature of the active events in various active gel systems. Our results may serve as motivation for detailed experimental  studies of the dynamics in active systems, artificial and biological, with strong elastic restoring forces.


\begin{acknowledgments}
We thank Yair Shokef, Yael Roichman, Anne Bernheim and Stanislav Burov for useful discussions. NSG is the incumbent of the Lee and William Abramowitz Professorial Chair of Biophysics and this research was made possible in part by the generosity of the Harold Perlman family. NSG acknowledges support from the ISF (Grant No. 580/12). NR acknowledges support from the Kimmelman Center for Biomolecular Structure and Assembly.
\end{acknowledgments}


\begin{thebibliography}{41}%
\makeatletter
\providecommand \@ifxundefined [1]{%
 \@ifx{#1\undefined}
}%
\providecommand \@ifnum [1]{%
 \ifnum #1\expandafter \@firstoftwo
 \else \expandafter \@secondoftwo
 \fi
}%
\providecommand \@ifx [1]{%
 \ifx #1\expandafter \@firstoftwo
 \else \expandafter \@secondoftwo
 \fi
}%
\providecommand \natexlab [1]{#1}%
\providecommand \enquote  [1]{``#1''}%
\providecommand \bibnamefont  [1]{#1}%
\providecommand \bibfnamefont [1]{#1}%
\providecommand \citenamefont [1]{#1}%
\providecommand \href@noop [0]{\@secondoftwo}%
\providecommand \href [0]{\begingroup \@sanitize@url \@href}%
\providecommand \@href[1]{\@@startlink{#1}\@@href}%
\providecommand \@@href[1]{\endgroup#1\@@endlink}%
\providecommand \@sanitize@url [0]{\catcode `\\12\catcode `\$12\catcode
  `\&12\catcode `\#12\catcode `\^12\catcode `\_12\catcode `\%12\relax}%
\providecommand \@@startlink[1]{}%
\providecommand \@@endlink[0]{}%
\providecommand \url  [0]{\begingroup\@sanitize@url \@url }%
\providecommand \@url [1]{\endgroup\@href {#1}{\urlprefix }}%
\providecommand \urlprefix  [0]{URL }%
\providecommand \Eprint [0]{\href }%
\providecommand \doibase [0]{http://dx.doi.org/}%
\providecommand \selectlanguage [0]{\@gobble}%
\providecommand \bibinfo  [0]{\@secondoftwo}%
\providecommand \bibfield  [0]{\@secondoftwo}%
\providecommand \translation [1]{[#1]}%
\providecommand \BibitemOpen [0]{}%
\providecommand \bibitemStop [0]{}%
\providecommand \bibitemNoStop [0]{.\EOS\space}%
\providecommand \EOS [0]{\spacefactor3000\relax}%
\providecommand \BibitemShut  [1]{\csname bibitem#1\endcsname}%
\let\auto@bib@innerbib\@empty
\bibitem [{\citenamefont {Backouche}\ \emph {et~al.}(2006)\citenamefont
  {Backouche}, \citenamefont {Haviv}, \citenamefont {Groswasser},\ and\
  \citenamefont {Bernheim-Groswasser}}]{Backouche2006}%
  \BibitemOpen
  \bibfield  {author} {\bibinfo {author} {\bibfnamefont {F.}~\bibnamefont
  {Backouche}}, \bibinfo {author} {\bibfnamefont {L.}~\bibnamefont {Haviv}},
  \bibinfo {author} {\bibfnamefont {D.}~\bibnamefont {Groswasser}}, \ and\
  \bibinfo {author} {\bibfnamefont {A.}~\bibnamefont {Bernheim-Groswasser}},\
  }\href {http://stacks.iop.org/1478-3975/3/i=4/a=004} {\bibfield  {journal}
  {\bibinfo  {journal} {Physical Biology}\ }\textbf {\bibinfo {volume} {3}},\
  \bibinfo {pages} {264} (\bibinfo {year} {2006})}\BibitemShut {NoStop}%
\bibitem [{\citenamefont {Mizuno}\ \emph {et~al.}(2007)\citenamefont {Mizuno},
  \citenamefont {Tardin}, \citenamefont {Schmidt},\ and\ \citenamefont
  {MacKintosh}}]{Mizuno2007}%
  \BibitemOpen
  \bibfield  {author} {\bibinfo {author} {\bibfnamefont {D.}~\bibnamefont
  {Mizuno}}, \bibinfo {author} {\bibfnamefont {C.}~\bibnamefont {Tardin}},
  \bibinfo {author} {\bibfnamefont {C.~F.}\ \bibnamefont {Schmidt}}, \ and\
  \bibinfo {author} {\bibfnamefont {F.~C.}\ \bibnamefont {MacKintosh}},\ }\href
  {\doibase 10.1126/science.1134404} {\bibfield  {journal} {\bibinfo  {journal}
  {Science}\ }\textbf {\bibinfo {volume} {315}},\ \bibinfo {pages} {370}
  (\bibinfo {year} {2007})}\BibitemShut {NoStop}%
\bibitem [{\citenamefont {Cates}\ \emph {et~al.}(2008)\citenamefont {Cates},
  \citenamefont {Fielding}, \citenamefont {Marenduzzo}, \citenamefont
  {Orlandini},\ and\ \citenamefont {Yeomans}}]{Cates2008}%
  \BibitemOpen
  \bibfield  {author} {\bibinfo {author} {\bibfnamefont {M.~E.}\ \bibnamefont
  {Cates}}, \bibinfo {author} {\bibfnamefont {S.~M.}\ \bibnamefont {Fielding}},
  \bibinfo {author} {\bibfnamefont {D.}~\bibnamefont {Marenduzzo}}, \bibinfo
  {author} {\bibfnamefont {E.}~\bibnamefont {Orlandini}}, \ and\ \bibinfo
  {author} {\bibfnamefont {J.~M.}\ \bibnamefont {Yeomans}},\ }\href {\doibase
  10.1103/PhysRevLett.101.068102} {\bibfield  {journal} {\bibinfo  {journal}
  {Phys. Rev. Lett.}\ }\textbf {\bibinfo {volume} {101}},\ \bibinfo {pages}
  {068102} (\bibinfo {year} {2008})}\BibitemShut {NoStop}%
\bibitem [{\citenamefont {MacKintosh}\ and\ \citenamefont
  {Levine}(2008)}]{MacKintosh2008}%
  \BibitemOpen
  \bibfield  {author} {\bibinfo {author} {\bibfnamefont {F.~C.}\ \bibnamefont
  {MacKintosh}}\ and\ \bibinfo {author} {\bibfnamefont {A.~J.}\ \bibnamefont
  {Levine}},\ }\href {\doibase 10.1103/PhysRevLett.100.018104} {\bibfield
  {journal} {\bibinfo  {journal} {Phys. Rev. Lett.}\ }\textbf {\bibinfo
  {volume} {100}},\ \bibinfo {pages} {018104} (\bibinfo {year}
  {2008})}\BibitemShut {NoStop}%
\bibitem [{\citenamefont {MacKintosh}\ and\ \citenamefont
  {Schmidt}(2010)}]{MacKintosh2010}%
  \BibitemOpen
  \bibfield  {author} {\bibinfo {author} {\bibfnamefont {F.~C.}\ \bibnamefont
  {MacKintosh}}\ and\ \bibinfo {author} {\bibfnamefont {C.~F.}\ \bibnamefont
  {Schmidt}},\ }\href {\doibase https://doi.org/10.1016/j.ceb.2010.01.002}
  {\bibfield  {journal} {\bibinfo  {journal} {Current Opinion in Cell Biology}\
  }\textbf {\bibinfo {volume} {22}},\ \bibinfo {pages} {29 } (\bibinfo {year}
  {2010})}\BibitemShut {NoStop}%
\bibitem [{\citenamefont {Toyota}\ \emph {et~al.}(2011)\citenamefont {Toyota},
  \citenamefont {Head}, \citenamefont {Schmidt},\ and\ \citenamefont
  {Mizuno}}]{Toyota2011}%
  \BibitemOpen
  \bibfield  {author} {\bibinfo {author} {\bibfnamefont {T.}~\bibnamefont
  {Toyota}}, \bibinfo {author} {\bibfnamefont {D.~A.}\ \bibnamefont {Head}},
  \bibinfo {author} {\bibfnamefont {C.~F.}\ \bibnamefont {Schmidt}}, \ and\
  \bibinfo {author} {\bibfnamefont {D.}~\bibnamefont {Mizuno}},\ }\href
  {\doibase 10.1039/C0SM00925C} {\bibfield  {journal} {\bibinfo  {journal}
  {Soft Matter}\ }\textbf {\bibinfo {volume} {7}},\ \bibinfo {pages} {3234}
  (\bibinfo {year} {2011})}\BibitemShut {NoStop}%
\bibitem [{\citenamefont {Stuhrmann}\ \emph {et~al.}(2012)\citenamefont
  {Stuhrmann}, \citenamefont {Soares~e Silva}, \citenamefont {Depken},
  \citenamefont {MacKintosh},\ and\ \citenamefont
  {Koenderink}}]{Stuhrmann2012}%
  \BibitemOpen
  \bibfield  {author} {\bibinfo {author} {\bibfnamefont {B.}~\bibnamefont
  {Stuhrmann}}, \bibinfo {author} {\bibfnamefont {M.}~\bibnamefont {Soares~e
  Silva}}, \bibinfo {author} {\bibfnamefont {M.}~\bibnamefont {Depken}},
  \bibinfo {author} {\bibfnamefont {F.~C.}\ \bibnamefont {MacKintosh}}, \ and\
  \bibinfo {author} {\bibfnamefont {G.~H.}\ \bibnamefont {Koenderink}},\ }\href
  {\doibase 10.1103/PhysRevE.86.020901} {\bibfield  {journal} {\bibinfo
  {journal} {Phys. Rev. E}\ }\textbf {\bibinfo {volume} {86}},\ \bibinfo
  {pages} {020901} (\bibinfo {year} {2012})}\BibitemShut {NoStop}%
\bibitem [{\citenamefont {e~Silva}\ \emph {et~al.}(2014)\citenamefont
  {e~Silva}, \citenamefont {Stuhrmann}, \citenamefont {Betz},\ and\
  \citenamefont {Koenderink}}]{Silva2014}%
  \BibitemOpen
  \bibfield  {author} {\bibinfo {author} {\bibfnamefont {M.~S.}\ \bibnamefont
  {e~Silva}}, \bibinfo {author} {\bibfnamefont {B.}~\bibnamefont {Stuhrmann}},
  \bibinfo {author} {\bibfnamefont {T.}~\bibnamefont {Betz}}, \ and\ \bibinfo
  {author} {\bibfnamefont {G.~H.}\ \bibnamefont {Koenderink}},\ }\href
  {http://stacks.iop.org/1367-2630/16/i=7/a=075010} {\bibfield  {journal}
  {\bibinfo  {journal} {New Journal of Physics}\ }\textbf {\bibinfo {volume}
  {16}},\ \bibinfo {pages} {075010} (\bibinfo {year} {2014})}\BibitemShut
  {NoStop}%
\bibitem [{\citenamefont {Sonn-Segev}\ \emph
  {et~al.}(2017{\natexlab{a}})\citenamefont {Sonn-Segev}, \citenamefont
  {Bernheim-Groswasser},\ and\ \citenamefont {Roichman}}]{Sonn-Segev2017}%
  \BibitemOpen
  \bibfield  {author} {\bibinfo {author} {\bibfnamefont {A.}~\bibnamefont
  {Sonn-Segev}}, \bibinfo {author} {\bibfnamefont {A.}~\bibnamefont
  {Bernheim-Groswasser}}, \ and\ \bibinfo {author} {\bibfnamefont
  {Y.}~\bibnamefont {Roichman}},\ }\href {\doibase 10.1039/C7SM01391D}
  {\bibfield  {journal} {\bibinfo  {journal} {Soft Matter}\ }\textbf {\bibinfo
  {volume} {13}},\ \bibinfo {pages} {7352} (\bibinfo {year}
  {2017}{\natexlab{a}})}\BibitemShut {NoStop}%
\bibitem [{\citenamefont {Sonn-Segev}\ \emph
  {et~al.}(2017{\natexlab{b}})\citenamefont {Sonn-Segev}, \citenamefont
  {Bernheim-Groswasser},\ and\ \citenamefont {Roichman}}]{Sonn-Segev2017JPCM}%
  \BibitemOpen
  \bibfield  {author} {\bibinfo {author} {\bibfnamefont {A.}~\bibnamefont
  {Sonn-Segev}}, \bibinfo {author} {\bibfnamefont {A.}~\bibnamefont
  {Bernheim-Groswasser}}, \ and\ \bibinfo {author} {\bibfnamefont
  {Y.}~\bibnamefont {Roichman}},\ }\href
  {http://stacks.iop.org/0953-8984/29/i=16/a=163002} {\bibfield  {journal}
  {\bibinfo  {journal} {Journal of Physics: Condensed Matter}\ }\textbf
  {\bibinfo {volume} {29}},\ \bibinfo {pages} {163002} (\bibinfo {year}
  {2017}{\natexlab{b}})}\BibitemShut {NoStop}%
\bibitem [{\citenamefont {Alvarado}\ \emph {et~al.}(2013)\citenamefont
  {Alvarado}, \citenamefont {Sheinman}, \citenamefont {Sharma}, \citenamefont
  {MacKintosh},\ and\ \citenamefont {Koenderink}}]{Alvarado2013}%
  \BibitemOpen
  \bibfield  {author} {\bibinfo {author} {\bibfnamefont {J.}~\bibnamefont
  {Alvarado}}, \bibinfo {author} {\bibfnamefont {M.}~\bibnamefont {Sheinman}},
  \bibinfo {author} {\bibfnamefont {A.}~\bibnamefont {Sharma}}, \bibinfo
  {author} {\bibfnamefont {F.~C.}\ \bibnamefont {MacKintosh}}, \ and\ \bibinfo
  {author} {\bibfnamefont {G.~H.}\ \bibnamefont {Koenderink}},\ }\href
  {http://dx.doi.org/10.1038/nphys2715} {\bibfield  {journal} {\bibinfo
  {journal} {Nature Physics}\ }\textbf {\bibinfo {volume} {9}},\ \bibinfo
  {pages} {591 EP } (\bibinfo {year} {2013})},\ \bibinfo {note}
  {article}\BibitemShut {NoStop}%
\bibitem [{\citenamefont {Marchetti}\ \emph {et~al.}(2013)\citenamefont
  {Marchetti}, \citenamefont {Joanny}, \citenamefont {Ramaswamy}, \citenamefont
  {Liverpool}, \citenamefont {Prost}, \citenamefont {Rao},\ and\ \citenamefont
  {Simha}}]{Marchetti2013RMP}%
  \BibitemOpen
  \bibfield  {author} {\bibinfo {author} {\bibfnamefont {M.~C.}\ \bibnamefont
  {Marchetti}}, \bibinfo {author} {\bibfnamefont {J.~F.}\ \bibnamefont
  {Joanny}}, \bibinfo {author} {\bibfnamefont {S.}~\bibnamefont {Ramaswamy}},
  \bibinfo {author} {\bibfnamefont {T.~B.}\ \bibnamefont {Liverpool}}, \bibinfo
  {author} {\bibfnamefont {J.}~\bibnamefont {Prost}}, \bibinfo {author}
  {\bibfnamefont {M.}~\bibnamefont {Rao}}, \ and\ \bibinfo {author}
  {\bibfnamefont {R.~A.}\ \bibnamefont {Simha}},\ }\href {\doibase
  10.1103/RevModPhys.85.1143} {\bibfield  {journal} {\bibinfo  {journal} {Rev.
  Mod. Phys.}\ }\textbf {\bibinfo {volume} {85}},\ \bibinfo {pages} {1143}
  (\bibinfo {year} {2013})}\BibitemShut {NoStop}%
\bibitem [{\citenamefont {Ben-Isaac}\ \emph {et~al.}(2015)\citenamefont
  {Ben-Isaac}, \citenamefont {Fodor}, \citenamefont {Visco}, \citenamefont {van
  Wijland},\ and\ \citenamefont {Gov}}]{BenIsaac2015}%
  \BibitemOpen
  \bibfield  {author} {\bibinfo {author} {\bibfnamefont {E.}~\bibnamefont
  {Ben-Isaac}}, \bibinfo {author} {\bibfnamefont {E.}~\bibnamefont {Fodor}},
  \bibinfo {author} {\bibfnamefont {P.}~\bibnamefont {Visco}}, \bibinfo
  {author} {\bibfnamefont {F.}~\bibnamefont {van Wijland}}, \ and\ \bibinfo
  {author} {\bibfnamefont {N.~S.}\ \bibnamefont {Gov}},\ }\href {\doibase
  10.1103/PhysRevE.92.012716} {\bibfield  {journal} {\bibinfo  {journal} {Phys.
  Rev. E}\ }\textbf {\bibinfo {volume} {92}},\ \bibinfo {pages} {012716}
  (\bibinfo {year} {2015})}\BibitemShut {NoStop}%
\bibitem [{\citenamefont {Klumpp}\ and\ \citenamefont
  {Lipowsky}(2005)}]{Klumpp2005}%
  \BibitemOpen
  \bibfield  {author} {\bibinfo {author} {\bibfnamefont {S.}~\bibnamefont
  {Klumpp}}\ and\ \bibinfo {author} {\bibfnamefont {R.}~\bibnamefont
  {Lipowsky}},\ }\href {\doibase 10.1073/pnas.0507363102} {\bibfield  {journal}
  {\bibinfo  {journal} {Proceedings of the National Academy of Sciences}\
  }\textbf {\bibinfo {volume} {102}},\ \bibinfo {pages} {17284} (\bibinfo
  {year} {2005})}\BibitemShut {NoStop}%
\bibitem [{\citenamefont {Gu\'erin}\ \emph {et~al.}(2010)\citenamefont
  {Gu\'erin}, \citenamefont {Prost}, \citenamefont {Martin},\ and\
  \citenamefont {Joanny}}]{Guerin2010}%
  \BibitemOpen
  \bibfield  {author} {\bibinfo {author} {\bibfnamefont {T.}~\bibnamefont
  {Gu\'erin}}, \bibinfo {author} {\bibfnamefont {J.}~\bibnamefont {Prost}},
  \bibinfo {author} {\bibfnamefont {P.}~\bibnamefont {Martin}}, \ and\ \bibinfo
  {author} {\bibfnamefont {J.-F.}\ \bibnamefont {Joanny}},\ }\href {\doibase
  https://doi.org/10.1016/j.ceb.2009.12.012} {\bibfield  {journal} {\bibinfo
  {journal} {Current Opinion in Cell Biology}\ }\textbf {\bibinfo {volume}
  {22}},\ \bibinfo {pages} {14 } (\bibinfo {year} {2010})},\ \bibinfo {note}
  {cell structure and dynamics}\BibitemShut {NoStop}%
\bibitem [{\citenamefont {Wang}\ and\ \citenamefont
  {Wolynes}(2011)}]{Wang2011}%
  \BibitemOpen
  \bibfield  {author} {\bibinfo {author} {\bibfnamefont {S.}~\bibnamefont
  {Wang}}\ and\ \bibinfo {author} {\bibfnamefont {P.~G.}\ \bibnamefont
  {Wolynes}},\ }\href {\doibase 10.1073/pnas.1112034108} {\bibfield  {journal}
  {\bibinfo  {journal} {Proceedings of the National Academy of Sciences}\
  }\textbf {\bibinfo {volume} {108}},\ \bibinfo {pages} {15184} (\bibinfo
  {year} {2011})}\BibitemShut {NoStop}%
\bibitem [{\citenamefont {Wang}\ and\ \citenamefont
  {Wolynes}(2012)}]{Wang2012}%
  \BibitemOpen
  \bibfield  {author} {\bibinfo {author} {\bibfnamefont {S.}~\bibnamefont
  {Wang}}\ and\ \bibinfo {author} {\bibfnamefont {P.~G.}\ \bibnamefont
  {Wolynes}},\ }\href {\doibase 10.1073/pnas.1204205109} {\bibfield  {journal}
  {\bibinfo  {journal} {Proceedings of the National Academy of Sciences}\
  }\textbf {\bibinfo {volume} {109}},\ \bibinfo {pages} {6446} (\bibinfo {year}
  {2012})}\BibitemShut {NoStop}%
\bibitem [{\citenamefont {Ucar}\ and\ \citenamefont
  {Lipowsky}(2017)}]{Ucar2017}%
  \BibitemOpen
  \bibfield  {author} {\bibinfo {author} {\bibfnamefont {M.~C.}\ \bibnamefont
  {Ucar}}\ and\ \bibinfo {author} {\bibfnamefont {R.}~\bibnamefont
  {Lipowsky}},\ }\href {\doibase 10.1039/C6SM01853J} {\bibfield  {journal}
  {\bibinfo  {journal} {Soft Matter}\ }\textbf {\bibinfo {volume} {13}},\
  \bibinfo {pages} {328} (\bibinfo {year} {2017})}\BibitemShut {NoStop}%
\bibitem [{\citenamefont {Sung}\ \emph {et~al.}(2015)\citenamefont {Sung},
  \citenamefont {Nag}, \citenamefont {Mortensen}, \citenamefont {Vestergaard},
  \citenamefont {Sutton}, \citenamefont {Ruppel}, \citenamefont {Flyvbjerg},\
  and\ \citenamefont {Spudich}}]{Sung2015}%
  \BibitemOpen
  \bibfield  {author} {\bibinfo {author} {\bibfnamefont {J.}~\bibnamefont
  {Sung}}, \bibinfo {author} {\bibfnamefont {S.}~\bibnamefont {Nag}}, \bibinfo
  {author} {\bibfnamefont {K.~I.}\ \bibnamefont {Mortensen}}, \bibinfo {author}
  {\bibfnamefont {C.~L.}\ \bibnamefont {Vestergaard}}, \bibinfo {author}
  {\bibfnamefont {S.}~\bibnamefont {Sutton}}, \bibinfo {author} {\bibfnamefont
  {K.}~\bibnamefont {Ruppel}}, \bibinfo {author} {\bibfnamefont
  {H.}~\bibnamefont {Flyvbjerg}}, \ and\ \bibinfo {author} {\bibfnamefont
  {J.~A.}\ \bibnamefont {Spudich}},\ }\href
  {http://dx.doi.org/10.1038/ncomms8931} {\bibfield  {journal} {\bibinfo
  {journal} {Nature Communications}\ }\textbf {\bibinfo {volume} {6}},\
  \bibinfo {pages} {7931} (\bibinfo {year} {2015})},\ \bibinfo {note}
  {article}\BibitemShut {NoStop}%
\bibitem [{\citenamefont {Gov}(2004)}]{Gov2004}%
  \BibitemOpen
  \bibfield  {author} {\bibinfo {author} {\bibfnamefont {N.}~\bibnamefont
  {Gov}},\ }\href {\doibase 10.1103/PhysRevLett.93.268104} {\bibfield
  {journal} {\bibinfo  {journal} {Phys. Rev. Lett.}\ }\textbf {\bibinfo
  {volume} {93}},\ \bibinfo {pages} {268104} (\bibinfo {year}
  {2004})}\BibitemShut {NoStop}%
\bibitem [{\citenamefont {Girard}\ \emph {et~al.}(2005)\citenamefont {Girard},
  \citenamefont {Prost},\ and\ \citenamefont {Bassereau}}]{Girard2005}%
  \BibitemOpen
  \bibfield  {author} {\bibinfo {author} {\bibfnamefont {P.}~\bibnamefont
  {Girard}}, \bibinfo {author} {\bibfnamefont {J.}~\bibnamefont {Prost}}, \
  and\ \bibinfo {author} {\bibfnamefont {P.}~\bibnamefont {Bassereau}},\ }\href
  {\doibase 10.1103/PhysRevLett.94.088102} {\bibfield  {journal} {\bibinfo
  {journal} {Phys. Rev. Lett.}\ }\textbf {\bibinfo {volume} {94}},\ \bibinfo
  {pages} {088102} (\bibinfo {year} {2005})}\BibitemShut {NoStop}%
\bibitem [{\citenamefont {{Turlier}}\ and\ \citenamefont
  {{Betz}}(2018)}]{Turlier2018}%
  \BibitemOpen
  \bibfield  {author} {\bibinfo {author} {\bibfnamefont {H.}~\bibnamefont
  {{Turlier}}}\ and\ \bibinfo {author} {\bibfnamefont {T.}~\bibnamefont
  {{Betz}}},\ }\href@noop {} {\bibfield  {journal} {\bibinfo  {journal} {ArXiv
  e-prints}\ } (\bibinfo {year} {2018})},\ \Eprint
  {http://arxiv.org/abs/1801.00176} {arXiv:1801.00176 [physics.bio-ph]}
  \BibitemShut {NoStop}%
\bibitem [{\citenamefont {Ben-Isaac}\ \emph {et~al.}(2011)\citenamefont
  {Ben-Isaac}, \citenamefont {Park}, \citenamefont {Popescu}, \citenamefont
  {Brown}, \citenamefont {Gov},\ and\ \citenamefont {Shokef}}]{Ben-Isaac2011}%
  \BibitemOpen
  \bibfield  {author} {\bibinfo {author} {\bibfnamefont {E.}~\bibnamefont
  {Ben-Isaac}}, \bibinfo {author} {\bibfnamefont {Y.}~\bibnamefont {Park}},
  \bibinfo {author} {\bibfnamefont {G.}~\bibnamefont {Popescu}}, \bibinfo
  {author} {\bibfnamefont {F.~L.~H.}\ \bibnamefont {Brown}}, \bibinfo {author}
  {\bibfnamefont {N.~S.}\ \bibnamefont {Gov}}, \ and\ \bibinfo {author}
  {\bibfnamefont {Y.}~\bibnamefont {Shokef}},\ }\href {\doibase
  10.1103/PhysRevLett.106.238103} {\bibfield  {journal} {\bibinfo  {journal}
  {Phys. Rev. Lett.}\ }\textbf {\bibinfo {volume} {106}},\ \bibinfo {pages}
  {238103} (\bibinfo {year} {2011})}\BibitemShut {NoStop}%
\bibitem [{\citenamefont {Loi}\ \emph {et~al.}(2011{\natexlab{a}})\citenamefont
  {Loi}, \citenamefont {Mossa},\ and\ \citenamefont {Cugliandolo}}]{Loi2011}%
  \BibitemOpen
  \bibfield  {author} {\bibinfo {author} {\bibfnamefont {D.}~\bibnamefont
  {Loi}}, \bibinfo {author} {\bibfnamefont {S.}~\bibnamefont {Mossa}}, \ and\
  \bibinfo {author} {\bibfnamefont {L.~F.}\ \bibnamefont {Cugliandolo}},\
  }\href {\doibase 10.1039/C1SM05819C} {\bibfield  {journal} {\bibinfo
  {journal} {Soft Matter}\ }\textbf {\bibinfo {volume} {7}},\ \bibinfo {pages}
  {10193} (\bibinfo {year} {2011}{\natexlab{a}})}\BibitemShut {NoStop}%
\bibitem [{\citenamefont {Ghosh}\ and\ \citenamefont {Gov}(2014)}]{Ghosh2014}%
  \BibitemOpen
  \bibfield  {author} {\bibinfo {author} {\bibfnamefont {A.}~\bibnamefont
  {Ghosh}}\ and\ \bibinfo {author} {\bibfnamefont {N.}~\bibnamefont {Gov}},\
  }\href {\doibase https://doi.org/10.1016/j.bpj.2014.07.034} {\bibfield
  {journal} {\bibinfo  {journal} {Biophysical Journal}\ }\textbf {\bibinfo
  {volume} {107}},\ \bibinfo {pages} {1065 } (\bibinfo {year}
  {2014})}\BibitemShut {NoStop}%
\bibitem [{\citenamefont {Weber}\ \emph {et~al.}(2015)\citenamefont {Weber},
  \citenamefont {Suzuki}, \citenamefont {Schaller}, \citenamefont {Aranson},
  \citenamefont {Bausch},\ and\ \citenamefont {Frey}}]{Weber2015}%
  \BibitemOpen
  \bibfield  {author} {\bibinfo {author} {\bibfnamefont {C.~A.}\ \bibnamefont
  {Weber}}, \bibinfo {author} {\bibfnamefont {R.}~\bibnamefont {Suzuki}},
  \bibinfo {author} {\bibfnamefont {V.}~\bibnamefont {Schaller}}, \bibinfo
  {author} {\bibfnamefont {I.~S.}\ \bibnamefont {Aranson}}, \bibinfo {author}
  {\bibfnamefont {A.~R.}\ \bibnamefont {Bausch}}, \ and\ \bibinfo {author}
  {\bibfnamefont {E.}~\bibnamefont {Frey}},\ }\href {\doibase
  10.1073/pnas.1421322112} {\bibfield  {journal} {\bibinfo  {journal}
  {Proceedings of the National Academy of Sciences}\ }\textbf {\bibinfo
  {volume} {112}},\ \bibinfo {pages} {10703} (\bibinfo {year} {2015})},\
  \Eprint
  {http://arxiv.org/abs/http://www.pnas.org/content/112/34/10703.full.pdf}
  {http://www.pnas.org/content/112/34/10703.full.pdf} \BibitemShut {NoStop}%
\bibitem [{\citenamefont {Gomez-Solano}\ \emph {et~al.}(2010)\citenamefont
  {Gomez-Solano}, \citenamefont {Bellon}, \citenamefont {Petrosyan},\ and\
  \citenamefont {Ciliberto}}]{Gomez-Solano2010}%
  \BibitemOpen
  \bibfield  {author} {\bibinfo {author} {\bibfnamefont {J.~R.}\ \bibnamefont
  {Gomez-Solano}}, \bibinfo {author} {\bibfnamefont {L.}~\bibnamefont
  {Bellon}}, \bibinfo {author} {\bibfnamefont {A.}~\bibnamefont {Petrosyan}}, \
  and\ \bibinfo {author} {\bibfnamefont {S.}~\bibnamefont {Ciliberto}},\ }\href
  {http://stacks.iop.org/0295-5075/89/i=6/a=60003} {\bibfield  {journal}
  {\bibinfo  {journal} {EPL (Europhysics Letters)}\ }\textbf {\bibinfo {volume}
  {89}},\ \bibinfo {pages} {60003} (\bibinfo {year} {2010})}\BibitemShut
  {NoStop}%
\bibitem [{\citenamefont {Berthier}\ and\ \citenamefont
  {Barrat}(2002)}]{Berthier2002}%
  \BibitemOpen
  \bibfield  {author} {\bibinfo {author} {\bibfnamefont {L.}~\bibnamefont
  {Berthier}}\ and\ \bibinfo {author} {\bibfnamefont {J.-L.}\ \bibnamefont
  {Barrat}},\ }\href {\doibase 10.1103/PhysRevLett.89.095702} {\bibfield
  {journal} {\bibinfo  {journal} {Phys. Rev. Lett.}\ }\textbf {\bibinfo
  {volume} {89}},\ \bibinfo {pages} {095702} (\bibinfo {year}
  {2002})}\BibitemShut {NoStop}%
\bibitem [{\citenamefont {Loi}\ \emph {et~al.}(2008)\citenamefont {Loi},
  \citenamefont {Mossa},\ and\ \citenamefont {Cugliandolo}}]{Loi2008}%
  \BibitemOpen
  \bibfield  {author} {\bibinfo {author} {\bibfnamefont {D.}~\bibnamefont
  {Loi}}, \bibinfo {author} {\bibfnamefont {S.}~\bibnamefont {Mossa}}, \ and\
  \bibinfo {author} {\bibfnamefont {L.~F.}\ \bibnamefont {Cugliandolo}},\
  }\href {\doibase 10.1103/PhysRevE.77.051111} {\bibfield  {journal} {\bibinfo
  {journal} {Phys. Rev. E}\ }\textbf {\bibinfo {volume} {77}},\ \bibinfo
  {pages} {051111} (\bibinfo {year} {2008})}\BibitemShut {NoStop}%
\bibitem [{\citenamefont {Loi}\ \emph {et~al.}(2011{\natexlab{b}})\citenamefont
  {Loi}, \citenamefont {Mossa},\ and\ \citenamefont {Cugliandolo}}]{Loi2011SM}%
  \BibitemOpen
  \bibfield  {author} {\bibinfo {author} {\bibfnamefont {D.}~\bibnamefont
  {Loi}}, \bibinfo {author} {\bibfnamefont {S.}~\bibnamefont {Mossa}}, \ and\
  \bibinfo {author} {\bibfnamefont {L.~F.}\ \bibnamefont {Cugliandolo}},\
  }\href {\doibase 10.1039/C0SM01484B} {\bibfield  {journal} {\bibinfo
  {journal} {Soft Matter}\ }\textbf {\bibinfo {volume} {7}},\ \bibinfo {pages}
  {3726} (\bibinfo {year} {2011}{\natexlab{b}})}\BibitemShut {NoStop}%
\bibitem [{\citenamefont {Wachsmuth}\ \emph {et~al.}(2000)\citenamefont
  {Wachsmuth}, \citenamefont {Waldeck},\ and\ \citenamefont
  {Langowski}}]{Malte2000}%
  \BibitemOpen
  \bibfield  {author} {\bibinfo {author} {\bibfnamefont {M.}~\bibnamefont
  {Wachsmuth}}, \bibinfo {author} {\bibfnamefont {W.}~\bibnamefont {Waldeck}},
  \ and\ \bibinfo {author} {\bibfnamefont {J.}~\bibnamefont {Langowski}},\
  }\href {\doibase https://doi.org/10.1006/jmbi.2000.3692} {\bibfield
  {journal} {\bibinfo  {journal} {Journal of Molecular Biology}\ }\textbf
  {\bibinfo {volume} {298}},\ \bibinfo {pages} {677 } (\bibinfo {year}
  {2000})}\BibitemShut {NoStop}%
\bibitem [{\citenamefont {Weber}\ \emph {et~al.}(2012)\citenamefont {Weber},
  \citenamefont {Spakowitz},\ and\ \citenamefont {Theriot}}]{Weber2012}%
  \BibitemOpen
  \bibfield  {author} {\bibinfo {author} {\bibfnamefont {S.~C.}\ \bibnamefont
  {Weber}}, \bibinfo {author} {\bibfnamefont {A.~J.}\ \bibnamefont
  {Spakowitz}}, \ and\ \bibinfo {author} {\bibfnamefont {J.~A.}\ \bibnamefont
  {Theriot}},\ }\href {\doibase 10.1073/pnas.1119505109} {\bibfield  {journal}
  {\bibinfo  {journal} {Proceedings of the National Academy of Sciences}\
  }\textbf {\bibinfo {volume} {109}},\ \bibinfo {pages} {7338} (\bibinfo {year}
  {2012})},\ \Eprint
  {http://arxiv.org/abs/http://www.pnas.org/content/109/19/7338.full.pdf}
  {http://www.pnas.org/content/109/19/7338.full.pdf} \BibitemShut {NoStop}%
\bibitem [{\citenamefont {Bruinsma}\ \emph {et~al.}(2014)\citenamefont
  {Bruinsma}, \citenamefont {Grosberg}, \citenamefont {Rabin},\ and\
  \citenamefont {Zidovska}}]{Bruinsma2014}%
  \BibitemOpen
  \bibfield  {author} {\bibinfo {author} {\bibfnamefont {R.}~\bibnamefont
  {Bruinsma}}, \bibinfo {author} {\bibfnamefont {A.}~\bibnamefont {Grosberg}},
  \bibinfo {author} {\bibfnamefont {Y.}~\bibnamefont {Rabin}}, \ and\ \bibinfo
  {author} {\bibfnamefont {A.}~\bibnamefont {Zidovska}},\ }\href {\doibase
  https://doi.org/10.1016/j.bpj.2014.03.038} {\bibfield  {journal} {\bibinfo
  {journal} {Biophysical Journal}\ }\textbf {\bibinfo {volume} {106}},\
  \bibinfo {pages} {1871 } (\bibinfo {year} {2014})}\BibitemShut {NoStop}%
\bibitem [{\citenamefont {Zidovska}(2015)}]{Zidovska2015}%
  \BibitemOpen
  \bibfield  {author} {\bibinfo {author} {\bibfnamefont {A.}~\bibnamefont
  {Zidovska}},\ }\href {\doibase 10.1016/j.bpj.2014.11.2961} {\bibfield
  {journal} {\bibinfo  {journal} {Biophysical Journal}\ }\textbf {\bibinfo
  {volume} {108}},\ \bibinfo {pages} {540a} (\bibinfo {year}
  {2015})}\BibitemShut {NoStop}%
\bibitem [{\citenamefont {Zidovska}(2017)}]{Zidovska2017}%
  \BibitemOpen
  \bibfield  {author} {\bibinfo {author} {\bibfnamefont {A.}~\bibnamefont
  {Zidovska}},\ }\href {\doibase 10.1016/j.bpj.2016.11.999} {\bibfield
  {journal} {\bibinfo  {journal} {Biophysical Journal}\ }\textbf {\bibinfo
  {volume} {112}},\ \bibinfo {pages} {180a} (\bibinfo {year}
  {2017})}\BibitemShut {NoStop}%
\bibitem [{\citenamefont {Bursac}\ \emph {et~al.}(2005)\citenamefont {Bursac},
  \citenamefont {Lenormand}, \citenamefont {Fabry}, \citenamefont {Oliver},
  \citenamefont {Weitz}, \citenamefont {Viasnoff}, \citenamefont {Butler},\
  and\ \citenamefont {Fredberg}}]{Bursac2005}%
  \BibitemOpen
  \bibfield  {author} {\bibinfo {author} {\bibfnamefont {P.}~\bibnamefont
  {Bursac}}, \bibinfo {author} {\bibfnamefont {G.}~\bibnamefont {Lenormand}},
  \bibinfo {author} {\bibfnamefont {B.}~\bibnamefont {Fabry}}, \bibinfo
  {author} {\bibfnamefont {M.}~\bibnamefont {Oliver}}, \bibinfo {author}
  {\bibfnamefont {D.~A.}\ \bibnamefont {Weitz}}, \bibinfo {author}
  {\bibfnamefont {V.}~\bibnamefont {Viasnoff}}, \bibinfo {author}
  {\bibfnamefont {J.~P.}\ \bibnamefont {Butler}}, \ and\ \bibinfo {author}
  {\bibfnamefont {J.~J.}\ \bibnamefont {Fredberg}},\ }\href
  {http://dx.doi.org/10.1038/nmat1404} {\bibfield  {journal} {\bibinfo
  {journal} {Nature Materials}\ }\textbf {\bibinfo {volume} {4}},\ \bibinfo
  {pages} {557 EP } (\bibinfo {year} {2005})}\BibitemShut {NoStop}%
\bibitem [{\citenamefont {{Fodor, \'E.}}\ \emph {et~al.}(2015)\citenamefont
  {{Fodor, \'E.}}, \citenamefont {{Guo, M.}}, \citenamefont {{Gov, N. S.}},
  \citenamefont {{Visco, P.}}, \citenamefont {{Weitz, D. A.}},\ and\
  \citenamefont {{van Wijland, F.}}}]{Fodor2015}%
  \BibitemOpen
  \bibfield  {author} {\bibinfo {author} {\bibnamefont {{Fodor, \'E.}}},
  \bibinfo {author} {\bibnamefont {{Guo, M.}}}, \bibinfo {author} {\bibnamefont
  {{Gov, N. S.}}}, \bibinfo {author} {\bibnamefont {{Visco, P.}}}, \bibinfo
  {author} {\bibnamefont {{Weitz, D. A.}}}, \ and\ \bibinfo {author}
  {\bibnamefont {{van Wijland, F.}}},\ }\href {\doibase
  10.1209/0295-5075/110/48005} {\bibfield  {journal} {\bibinfo  {journal}
  {EPL}\ }\textbf {\bibinfo {volume} {110}},\ \bibinfo {pages} {48005}
  (\bibinfo {year} {2015})}\BibitemShut {NoStop}%
\bibitem [{\citenamefont {Fodor}\ \emph {et~al.}(2018)\citenamefont {Fodor},
  \citenamefont {Mehandia}, \citenamefont {Comelles}, \citenamefont
  {Thiagarajan}, \citenamefont {Gov}, \citenamefont {Visco}, \citenamefont {van
  Wijland},\ and\ \citenamefont {Riveline}}]{Fodor2018}%
  \BibitemOpen
  \bibfield  {author} {\bibinfo {author} {\bibfnamefont {{\'E}.}~\bibnamefont
  {Fodor}}, \bibinfo {author} {\bibfnamefont {V.}~\bibnamefont {Mehandia}},
  \bibinfo {author} {\bibfnamefont {J.}~\bibnamefont {Comelles}}, \bibinfo
  {author} {\bibfnamefont {R.}~\bibnamefont {Thiagarajan}}, \bibinfo {author}
  {\bibfnamefont {N.~S.}\ \bibnamefont {Gov}}, \bibinfo {author} {\bibfnamefont
  {P.}~\bibnamefont {Visco}}, \bibinfo {author} {\bibfnamefont
  {F.}~\bibnamefont {van Wijland}}, \ and\ \bibinfo {author} {\bibfnamefont
  {D.}~\bibnamefont {Riveline}},\ }\href {\doibase 10.1016/j.bpj.2017.12.026}
  {\bibfield  {journal} {\bibinfo  {journal} {Biophysical Journal}\ }\textbf
  {\bibinfo {volume} {114}},\ \bibinfo {pages} {939} (\bibinfo {year}
  {2018})}\BibitemShut {NoStop}%
\bibitem [{\citenamefont {Ahmed}\ \emph {et~al.}(2018)\citenamefont {Ahmed},
  \citenamefont {Fodor}, \citenamefont {Almonacid}, \citenamefont {Bussonnier},
  \citenamefont {Verlhac}, \citenamefont {Gov}, \citenamefont {Visco},
  \citenamefont {van Wijland},\ and\ \citenamefont {Betz}}]{Ahmed2018}%
  \BibitemOpen
  \bibfield  {author} {\bibinfo {author} {\bibfnamefont {W.~W.}\ \bibnamefont
  {Ahmed}}, \bibinfo {author} {\bibfnamefont {{\'E}.}~\bibnamefont {Fodor}},
  \bibinfo {author} {\bibfnamefont {M.}~\bibnamefont {Almonacid}}, \bibinfo
  {author} {\bibfnamefont {M.}~\bibnamefont {Bussonnier}}, \bibinfo {author}
  {\bibfnamefont {M.-H.}\ \bibnamefont {Verlhac}}, \bibinfo {author}
  {\bibfnamefont {N.}~\bibnamefont {Gov}}, \bibinfo {author} {\bibfnamefont
  {P.}~\bibnamefont {Visco}}, \bibinfo {author} {\bibfnamefont
  {F.}~\bibnamefont {van Wijland}}, \ and\ \bibinfo {author} {\bibfnamefont
  {T.}~\bibnamefont {Betz}},\ }\href {\doibase 10.1016/j.bpj.2018.02.009}
  {\bibfield  {journal} {\bibinfo  {journal} {Biophysical Journal}\ }\textbf
  {\bibinfo {volume} {114}},\ \bibinfo {pages} {1667} (\bibinfo {year}
  {2018})}\BibitemShut {NoStop}%
\bibitem [{Sup()}]{Supplemental}%
  \BibitemOpen
  \href@noop {} {}\bibinfo {note} {See Supplemental Material}\BibitemShut
  {NoStop}%
\end{thebibliography}
%

\end{document}